\documentclass[manuscript]{aastex}

\shorttitle{GrayStar II: Methods}
\shortauthors{Short}

\begin{document}


\title{GrayStar: A Web application for pedagogical stellar atmosphere and spectral line modelling 
and visualisation II: Methods}


\author{C. Ian Short}
\affil{Department of Astronomy \& Physics and Institute for Computational Astrophysics, Saint Mary's University,
    Halifax, NS, Canada, B3H 3C3}
\email{ishort@ap.smu.ca}




\begin{abstract}

GrayStar is a stellar atmospheric and spectral line modelling, post-processing, and visualisation code, suitable for 
classroom demonstrations and laboratory-style assignments, that has been developed in Java and deployed in 
JavaScript and HTML.  The only software needed to compute models and post-processed
observables, and to visualise the resulting
atmospheric structure and observables, is a common Web browser.  Therefore, the code will run on any common PC or related X86 (-64) computer 
of the type that typically serves classroom data projectors, is found in undergraduate computer laboratories, 
or that students themselves own, including those with highly portable form-factors such as net-books and tablets.  
The user requires no experience
with compiling source code, reading data files, or using plotting packages.  More advanced students can
view the JavaScript source code using the developer tools provided by common Web browsers. 
 The code is based 
on the approximate gray atmospheric solution and runs quickly enough on current common PCs to provide 
near-instantaneous results,
allowing for real time exploration of parameter space.  
I describe the
computational strategy and methodology as necessitated by Java and JavaScript.
In an accompanying paper, I describe the user interface and its inputs and outputs and suggest
specific pedagogical applications and projects.  
I have made the application itself, and the HTML, CSS, JavaScript,
and Java source files available to the community.  
The Web application and source files may be found at {\url www.ap.smu.ca/$\sim$ishort/GrayStar}.

\end{abstract}


\keywords{astronomy: education, stars: atmospheres, spectra}

\section{Introduction}

\paragraph{}

 GrayStar is a simple stellar atmospheric
 and spectral line modelling, post-processing, and visualisation code that has been designed to be suitable for  
pedagogical use by instructors and students with no experience with producing an executable 
file from source code, or with producing routines to read data files or with data visualisation, and who 
have access only to computers with mass-market GUI-based OSes.  The atmospheric structure is computed
 using the approximate gray solution (among other less crude approximations described below), obviating the need
 for input-output- (IO- ) intensive atomic and molecular ``big data'' handling and for iterative 
convergence.  The code is written in JavaScript, is processed by a Web browser's JavaScript interpreter and
the client's CPU, and 
displays its results in the browser window.  Therefore, it is certain to run successfully 
on any computer platform for which a common Web browser is available, 
 which includes all mass-market X86 (and X86-64) platforms and OSes of the type that serve classroom
 data projectors, are found in university computer labs, and that are owned by students and instructors,
 including those machines with highly portable form-factors such as tablets and net-books.  In \citet{GrayStarI}, 
hereafter Paper I, I describe the pedagogical context and present some pedagogical 
applications for GrayStar.  Here, I elaborate on the computational modelling strategy and numerical methods, 
especially as constrained by the peculiarities of Java and JavaScript, languages that are not traditionally
used in scientific programming.

\paragraph{}

In addition to making the GrayStar executable universally and freely available through the World Wide
 Web ({\url www.ap.smu.ca/$\sim$ishort/GrayStar}), I also disseminate the JavaScript and Hypertext Mark-up Language (HTML) source code,
and the source files for the Java {\it development} version (see below), to any who are interested in 
having their own local installation, or in developing the code further.  
I stress the broader significance that common PCs running common OSes are now powerful enough, and common Web-browsers are now sophisticated
enough interpreters, that the the realm of pedagogically useful scientific programming is now accessible in a framework that is free,
common, and allows both the application and its source code to be immediately shared over the Web.

\paragraph{}

In Section \ref{sUI} I review the GrayStar user interface (also described in Paper I); 
in Section \ref{sCode} I describe the modelling approach and computational methods, and remark on 
  special considerations relevant to developing and
deploying scientific modelling codes in Java and JavaScript; and in Section \ref{sApps} I
make brief comments on pedagogical aspects, although more detail of these is given in 
Paper I. 

\begin{figure}
\plotone{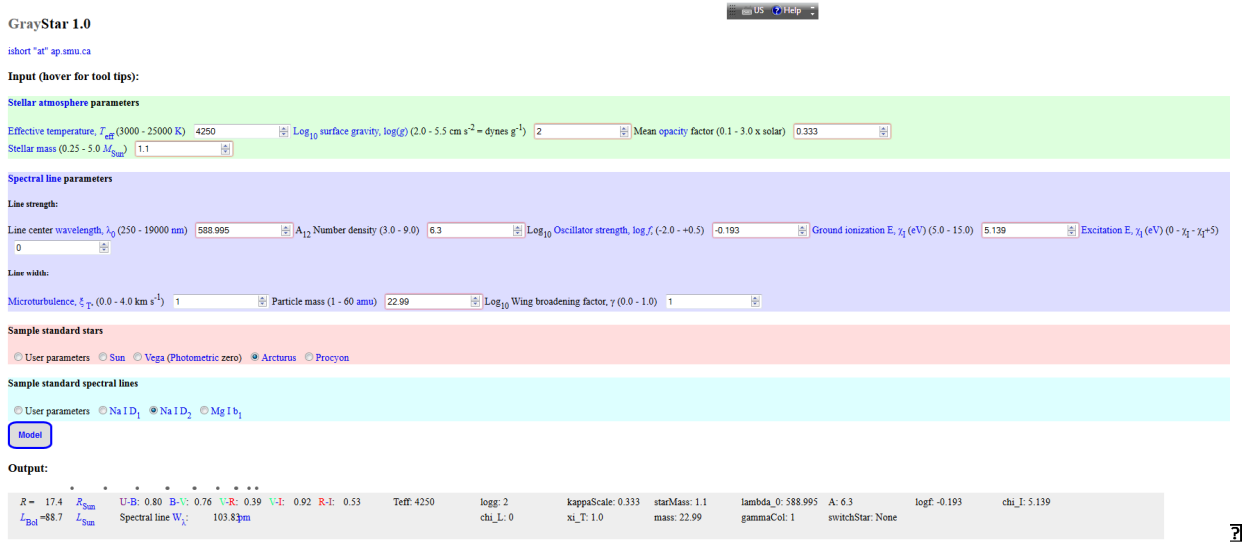}
\caption{A screen-shot of the input and textual output areas of the GrayStar GUI.
  \label{fGUIin}
}
\end{figure}

\begin{figure}
\plotone{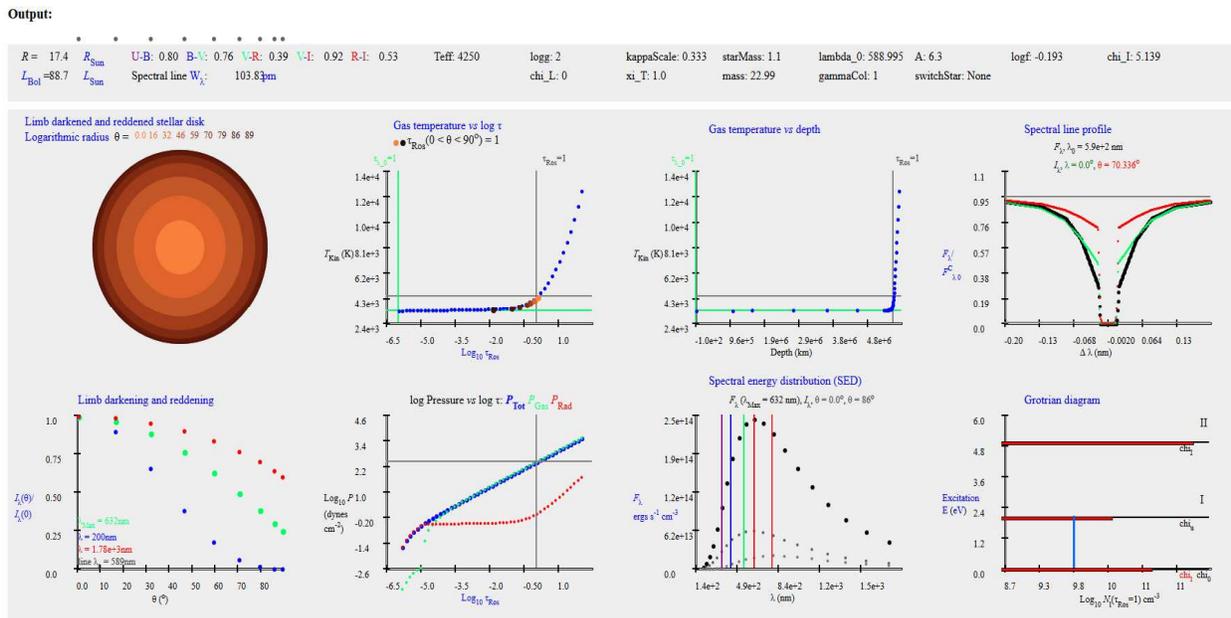}
\caption{A screen-shot of the output area of the GrayStar GUI showing the textual output section
and the eight plots of the graphical output section.
  \label{fGUIout}
}
\end{figure}


\section{User interface \label{sUI}}

\subsection{Input}

GrayStar presents the user with a browser window with 12 labeled 'text-box'-style
 input fields (Fig \ref{fGUIin}), four for stellar parameters and eight for spectral line parameters.
(Note that in
 the case of a gray atmospheric model, the microturbulence parameter, $\xi_{\rm T}$, is {\it only} a
 spectral line parameter, {\it not} an atmospheric parameter, because line extinction plays no role 
in determining the atmospheric structure.)  

\subsubsection{Stellar parameters }

1) Effective temperature, $T_{\rm Eff}$, in K, 2) Logarithmic surface gravity, $\log_{\rm 10} g$, 
in $\log$ cm s$^{-2}$ ($\log$ dynes g$^{-1}$), and 3) Multiplicative factor, $x$, for the Rosseland mean mass extinction
 co-efficient, $\kappa_{\rm Ros}$.  These are required to compute a model.  In addition, the interface expects 
4) Stellar mass, $M$, in solar units, $M_{\odot}$, for calculating the radius, $R$, and thus 
the bolometric luminosity, $L_{\rm Bol}$, in solar units, for purely pedagogical reasons.

\subsubsection{Spectral line parameters }

These are divided into two groups: those that mainly affect line {\it strength}, and those that mainly
affect line {\it width} or shape:

\paragraph{Line strength }
1) Line center wavelength, $\lambda_{\rm 0}$, in nm, 2) Logarithmic {\it total}
number density of the extinguishing species in the ``A$_{12}$'' system, $A = \log_{\rm 10} N/N_{\rm H} + 12$, 
3) Unit-less quantum
mechanical oscillator strength, $f_{\rm lu}$, of the corresponding bound-bound ({\it b-b}) atomic
transition, 4) Ground state ionization energy, $\chi_{\rm I}$, of the neutral ionization stage in eV, 5)
 Excitation energy, $\chi_{\rm l}$ of the lower atomic energy level of the corresponding {\it b-b}
transition, in eV, with respect to the ground state of the {\it neutral} stage.  If the value of
$\chi_{\rm l}$ (line parameter 5) is {\it less} than that of $\chi_{\rm I}$ (parameter 4), the spectral
 line corresponds to a {\it b-b} transition of the {\it neutral} ionization stage (I).  Otherwise, it
corresponds to a {\it b-b} transition of the {\it singly ionized} stage (II).

\paragraph{Line width and shape }

6) Mass of the absorbing species in atomic mass units (amu) (affects {\it thermal} core broadening), 
7) Microturbulent RMS velocity, $\xi_{\rm T}$, in km s$^{-1}$ (affects {\it non-thermal} core broadening),  
8) Logarithmic van der Waals damping enhancement factor, $\log_{\rm 10}\gamma$ (s$^{-1}$) 
(affects {\it wing} damping ).

\paragraph{}

Because this is a pedagogical application, guidance is provided for determining physically realistic
 and illustrative values.  The four stellar parameter input fields are pre-filled with default 
values for the Sun ($T_{\rm Eff}/\log g/x/M = 5778/4.44/1.0/1.0$), and the eight spectral line parameter 
fields are pre-filled with default values that yield a moderate spectral line ({\it ie.} with an 
approximately Gaussian profile) for a star with the Sun's stellar parameters.  The cgs system of units is used 
consistently, except for those values where standard practice is to use other units to ensure well-
normalized quantities (km s$^{-1}$ for $\xi_{\rm T}$, eV for atomic $\chi$ values).  
Moreover, the input fields are annotated with suggested 
ranges for the input values, and these ranges are enforced in the code itself to prevent students from
 inadvertently crashing the code by entering values that would lead to numerical pathologies.  I have
 also added ``tool tips'' to those field labels that are less self-explanatory that present additional information about the input 
parameters when the user hovers over a label, and most parameters labels are also linked to explanatory Web pages. 
If a user has their own installation of GrayStar, they can edit the links so that the fields point
to local on-line resources, thus embedding GrayStar in a local pedagogical framework.

\subsubsection{Pre-set models}
 
Finally, the user is given the option
of selecting from among four pre-set standard stars: the Sun (G2 V), Vega (A0 V, photometric zero), Arcturus (K1.5 III), 
and Procyon (F5 V-IV), and three pre-set Fraunhofer lines: The \ion{Na}{1} D$_{\rm 1}$ and D$_{\rm 2}$ lines, and the
 \ion{Mg}{1} b$_{\rm 1}$ line.  Comparison of the \ion{Na}{1} D$_{\rm 1}$ and D$_{\rm 2}$ lines demonstrates
COG effects in a multiplet (a doublet in this case), and the  \ion{Mg}{1} b$_{\rm 1}$ line demonstrates the
$T_{\rm Eff}$ variation of a line arising from an excited level.

\subsection{Output}

When the user runs a calculation by clicking the ``Model'' button, textual information and 
eight graphs are immediately displayed with the results of the calculation (Fig. \ref{fGUIout}):

\paragraph{Textual output: }

The values of the 12 input parameters are echoed back to the user.  This is important because 
the parameters used in the model may differ from those supplied if one or more parameters was
outside the ``safe'' range.  Any parameter that has been over-ridden is highlighted in red
to draw the student's attention.
The computed values of $R$ and $L_{\rm Bol}$ are displayed in solar units,  
along with five photometric colour indices in the Johnson-
Cousins  $UBV(RI)_{\rm C}$ system ($U-B$, $B-V$, $V-R$, $V-I$, $R-I$).  The colour indices are normalized 
with a single-point 
calibration to a GrayStar model computed with Vega's input parameters ($T_{\rm Eff}/log g/x = 9550/3.95/0.333$ \citep{castelli}).  
The equivalent width, $W_\lambda$, 
of the spectral line is displayed in pm (picometres).  The non-standard unit, pm (equal to 10 m\AA), was chosen for 
pedagogically-motivated consistency with the units of wavelength (nm).  

\paragraph{Graphical output: }

The GUI displays graphs of 
1) A physically based limb-darkened and limb-{\it reddened} rendering of the
projected stellar disk scaled logarithmically with radius, 
2) $T_{\rm Kin}$ {\it vs} logarithmic Rosseland optical depth, 
$\log\tau_{\rm Ros}$, 
3) Kinetic temperature, $T_{\rm Kin}$, in K {\it vs} 
geometric depth in km, 
4) logarithmic total pressure ($\log P_{\rm Tot}$), gas pressure 
($\log P_{\rm Gas}$), and bolometric radiation pressure ($\log P_{\rm Rad}$) {\it vs} 
$\log\tau_{\rm Ros}$, 
5) Limb darkening profiles, {$I_\lambda(\theta)\over I_\lambda(0)$ {\it vs} 
$\theta$ at the wavelength of maximum flux ($\lambda_{\rm Max}$), at representative continuum 
wavelengths in the near UV and near IR, and at spectral line center ($\lambda_{\rm 0}$), 
6) Surface flux 
spectral energy distribution (SED), $F_\lambda(\lambda)$ {\it vs} $\lambda$, and surface intensity 
SEDs, $I_\lambda(\lambda)$ at $\theta$ values of $\approx 0^\circ$ and $\approx 87^\circ$ for the 200
 (UV) to 2000 (IR) nm $\lambda$ range, 
7) the $F_\lambda(\lambda)$ spectral line profile and the
 $I_\lambda(\lambda)$ line profiles at  $\theta$ values of $\approx 60^\circ$ and $\approx 87^\circ$,
and 
8) A Grotrian diagram showing the atomic energy, $\chi_{\rm l}$, and logarithmic level population, $N_{\rm l}$,
at $\tau_{\rm Ros} = 1$ of four key $E$-levels: the lower and upper levels of the $b-b$ transition and the ground states of the
neutral and ionized stages, along with the transition itself.  

\paragraph{}

Plot 1) is labeled with the $\theta$ values of the emergent $I_\lambda(\theta)$ beams with respect
to the local surface-normal, color coded for consistency with the corresponding annuli in the 
limb-darkened, limb-reddened image. 
In plot 2) of $T_{\rm Kin}$ {\it vs} $\log\tau_{\rm Ros}$ the depths of $\tau_{\theta} = 1$ for
each $I_\lambda(\theta)$ beam sampling the radiation field in angle is shown with colour-coded symbols that
correspond to the limb- darkened and reddened rendering in plot 1).  Therefore, plots 1), 2), and 5) 
work together to provide a powerful, direct demostration of limb-darkening in terms of the LTE Eddington-Barbier relation.  
In the plots 2) through 4) displaying the atmospheric structure the depths where the continuum and line-center 
monochromatic optical depths scales are approximately unity (depths of $\tau_{\rm Ros}$ and $\tau_{\lambda 0} \approx 1.0$) 
are indicated, providing a powerful demostration of spectral line formation in terms of the LTE Eddington-Barbier relation.  
In plot 6) displaying the SEDs the central wavelengths of the 
$UBV(RI)_{\rm C}$ bands 
are indicated with appropriately colour-coded markers, and the value of ($\lambda_{\rm Max}$) is displayed. 

\section{Computational methods and approximations \label{sCode}}

\paragraph{Java and JavaFX development }

 To take advantage of the robust development support provided by Integrated Development Environment 
applications (IDEs), I developed the code in Java, using version 1.8 of the Java Development Kit (JDK 1.8).  
Java has strong data typing and interface declaration rules,
 allowing the IDE to immediately catch bugs caused by most common coding errors as the code is being 
developed, including those arising from a mis-match between the types and numbers of the arguments and 
of the parameters of a function ({\it ie.} a Java ``method'').  Additionally the IDE provides the usual 
standard out (stdout) and standard error (stderr) channels for the Java interpreter/compiler to communicate 
with the developer, and provides robust error messaging.  All these are crucial to developing scientific 
computational codes of even moderate complexity.  This {\it development} version of the code uses the 
JavaFX library to provide the GUI, and has thus been named GrayFox, and has also been made available to 
the community through the GrayStar WWW site.  

\paragraph{JavaScript and HTML deployment }

Because the Java Run-time Environment (JRE) allows 
pre-compiled executable code to be downloaded and run on the client, and because Java has full 
file I/O capability, it poses a significant security threat.  Therefore, a difficult and financially 
expensive security protocol requires the deployer of Java codes to digitally authenticate their code 
with a certificate purchased from a trusted certificate issuer, and this poses a significant barrier 
to the free and wide dissemination of such codes in the academic sector.
To circumvent this difficulty of Java deployment, the code was ported to JavaScript and HTML for 
Web deployment.  With JavaScript applications, the client browser down-loads source code that can be visually 
inspected with typical browser developer tools, and JavaScript does not have file I/O capability.  
Therefore, the onerous and expensive 
authentication protocol is not required, and the code can be executed transparently by a client with 
typical default security settings.  
However, because JavaScript does not have IDE support, nor file I/O capability, 
it is more difficult to trouble-shoot and debug.  
Therefore, the recommended 
work-flow for further development is to develop the code in Java, then port it to JavaScript and HTML 
for deployment.  Porting the modelling algorithms is straightforward because, with the exception of variable and function 
(Java method) declarations, the syntax is identical (an exception is object declaration, but I 
have been unable to think of a way in which object oriented programming would benefit atmospheric 
modelling and spectrum synthesis!).  This {\it deployment} version is called GrayStar.      
Because Java and JavaScript development of modelling codes is not a strong part of the scientific programming culture,
and because I will make the source code publicly available for those who wish to understand and develop it, or
have their students study and modify it,
I provide a significant level of detail below.  

\paragraph{HTML visualisation }

The biggest dis-incentive to scientific programming and visualisation with JavaScript and HTML is the need to
manually emulate the functionality of a plotting package ({\it egs.} IDL, gnuplot) starting from the primitive ability
of HTML to place a rectangle of given dimensions at a given location in the browser window, as specified in absolute device coordinates.
However, the code in the graphical output section of GrayStar may be taken as a template for how this problem can be
addressed, and adapted to other uses.

\paragraph{ }

GrayStar solves the static, $1D$ plane-parallel, horizontally homogeneous, local thermodynamic equilibrium 
(LTE), gray atmosphere problem, evaluates the formal solution of the radiative transfer equation to compute the
outgoing surface monochromatic specific intensity, $I_\lambda(\tau_\lambda=0, \theta)$,  and computes various
observables including the SED, photometric colour indices, and the profile of a representative spectral absorption line 
using the ``core plus wing'' approximation to a Voigt function profile.  (I note that all these approximations, 
{\it except the gray solution}, are not especially restrictive in the context of research-grade modelling!)  
The theoretical basis is taken 
entirely from \citet{rutten}.  The least realistic of these assumptions, by far, and the most expediting, 
is the gray solution, in which the monochromatic mass extinction co-efficient, $\kappa_\lambda(\lambda)$ 
is assumed to be constant as a function of wavelength, $\lambda$ (the Gray approximation), although it varies with depth,
and the angle-moments of the radiation field are related through the first and second Eddington approximations. 
 This yields an enormous simplification of the atmospheric structure problem in that the vertical kinetic 
temperature structure, $T_{\rm Kin}(\tau_{\rm Ros})$, can be calculated analytically, and the remaining structure variables 
({\it eg.} pressure ($P(\tau_{\rm Ros})$), density ($\rho(\tau_{\rm Ros})$)) can be calculated in a single pass with no need for iterative 
convergence.  The gray $T_{\rm Kin}(\tau)$ structure is most conceptually self-consistent at depth in the atmosphere when 
it is computed on the Rosseland optical depth scale, $\tau_{\rm Ros}$ ({\it ie.},  
$T_{\rm Kin}(\tau_{\rm Ros}))$.  Therefore, I set the gray value of $\kappa_\lambda$ at each depth to be equal (approximately) 
to the corresponding Rosseland mean mass extinction coefficient ({\it ie.} 
$\kappa_\lambda(\tau_{\rm Ros}, \lambda) = \kappa_{\rm Ros}(\tau_{\rm Ros})$) using the procedure
described below.

\paragraph{}
 Below I describe the calculations performed by GrayStar upon each execution, in the order of execution,
along with the approximations and methods employed.  The numbers of points sampling the atmosphere in
$\tau_{\rm Ros}$, and the radiation field in $\lambda$ and $\theta$, have been set to values close to
the minimum that are useful to optimize the speed of execution.
I employ the standard methods for minimizing the risk of under- and over- flows:
 All quantities are represented with intrinsic double precision data type (I note that although JavaScript
lacks explicit data typing, JavaScript enforces {\it double}-precision floating point variables in any case!).
Wherever possible,
the right-hand-side (RHS) of equations is evaluated logarithmically in base $e$ (natural logarithm, $\ln$),
and the result exponentiated as needed.  Differential equations are evaluated as first order finite
difference equations using Euler's method, and integrals are evaluated as finite sums with the extended
rectangle rule for computational expediency.  Tests comparing the results
with those computed with the fourth order Runge-Kutta (RK4) method and the extended Simpson's rule,
respectively, showed differences that were negligible for
pedagogical purposes.
Most atmospheric structure quantities are stored in $2D$ arrays
of dimension $2 \times N$, where $N$ is normally either the number of depth points or the number of
wavelength points, in which row $0$ of the array holds linear quantities, and row $1$ holds logarithmic
quantities.  This approach minimizes the number of logarithm and exponentiation operations needed throughout
the calculation, and is necessitated by the structure of Java, in which a {\it method} (equivalent to
a FORTRAN function) can only return a single data structure.  My testing indicates that the code is robust
for $T_{\rm Eff}$ values ranging from at least 3000 to 40000 K (MK spectral class M to O) and $\log g$
values ranging from at least 1.0 to 5.5 (MK luminosity class I (supergiants) to VI (sub-dwarfs)).

\paragraph{}

The following Java classes contain the methods that handle each module of the model atmosphere and spectral
energy distribution (SED) problem, and are contained in Java package grayfox in the development version (see above).
The main method, main(), from which the application is launched, is contained in Java class GrayFox (grayfox.GrayFox.main()).

\paragraph{}

1) Method tauScale() in Java class TauScale (grayfox.TauScale.tauScale(), ported to JavaScript function tauScale()) computes the values of 50
$\tau_{\rm Ros}$ points distributed evenly in $\log\tau_{\rm Ros}$ from -6.0 to 2.0.  This establishes an independent
variable for the depth scale, and is the same for all choices of atmospheric parameters.

\paragraph{}

2) Method temperature in Java class Temperature (grayfox.Temperature.temperature(), ported to JavaScript function temperature()) returns the value
of $T_{\rm Kin}$ at each of the $\tau_{\rm Ros}$ points by evaluating the gray formula for
$T_{\rm Kin}(\tau_{\rm Ros})$.  I use the more realistic variation of the formula with the Hopf function, $q(\tau_{\rm Ros})$,
with the value of $q(\tau_{\rm Ros})$ linearly interpolated in  $\log\tau_{\rm Ros}$ from 0.55 at
$\log\tau_{\rm Ros} = -6.0$ to 0.710 at $\log\tau_{\rm Ros} = 2.0$.

\paragraph{}

3) Method kappas() in Java class Kappas (grayfox.Kappas.kappas(), ported to JavaScript function kappas()) computes the logarithmic value of the
Rosseland mean mass extinction coefficient, $\log\kappa_{\rm Ros}$, in cm$^2$ g$^{-1}$, by linear interpolation
in $\log\tau_{\rm Ros}$, of the solar values taken from an Atlas9 \citet{kurucz} model of the Sun, and range from -3.5
$\log$ cm$^2$ g$^{-1}$ at $\log\tau_{\rm Ros} = -6.0$ to 2.0 $\log$ cm$^2$ g$^{-1}$ at $\log\tau_{\rm Ros} = 2.0$.
 These $\kappa_{\rm Ros}$  values are then re-scaled in two ways: a) They are multiplied by the value
of the input parameter, $x$, described in Section \ref{sUI};
b) For models of $T_{\rm Eff} > T_{\rm Hot}$, I add, depth-wise,
$2\times(\tau_{\rm Ros, Max} - \tau_{\rm Ros})/(\tau_{\rm Ros, Max} - \tau_{\rm Ros, Min})\times(T_{\rm Eff} - T_{\rm Hot})/T_{\rm Hot}$,
where $\tau_{\rm Ros, Max} = 2$ and $T_{\rm Hot} = 9500$ K,
to emulate the flattening of the  $\kappa_{\rm Ros}$ structure with increasing $T_{\rm Eff}$ caused by Thomson
scattering opacity from free electrons liberated by increasingly ionized \ion{H}{1}.
These $\log\kappa_{\rm Ros}$ values are inconsistent with the $\tau_{\rm 
Ros}$ values computed in Step 1).  However, because I am prescribing the $\tau_{\rm Ros}$ scale rather
than the geometric depth scale as the independent depth variable, I need an {\it ad hoc}
$\kappa_{\rm Ros}$ structure to integrate the equation of hydrostatic equilibrium (HSE).

\paragraph{}

4) Method hydrostatic() in Java class Hydrostat (grayfox.Hydrostat.hydrostatic(), ported to JavaScript function hydrostatic()) computes the total
pressure structure, $P(\log\tau_{\rm Ros})$, by numerically evaluating the first order finite difference for of
the hydrostatic equilibrium equation
(HSE) on the $\tau_{\rm Ros}$ scale with Euler's method (see note of numerical integration above). I adopt an {\it ad
hoc} upper boundary value of -4.0 for $\log P$ at $\log\tau_{\rm Ros} = -6.0$.
I then
compute the bolometric radiation pressure at each depth point, $P_{\rm Rad}(\log\tau_{\rm Ros})$, under
the assumption of black-body (BB) radiation, and subtract it from the $P(\log\tau_{\rm Ros})$ structure
to determine the gas pressure structure $P_{\rm Gas}(\log\tau_{\rm Ros})$ under the assumption that gas
and radiation are the only sources of pressure.  The subtraction is not evaluated directly, but rather
as $P_{\rm Gas} = P\times (1.0 - P_{\rm Rad}/P)$ to avoid loss of numerical precision when
$P_{\rm Rad}\approx P$ and both are large.  If $P_{\rm Rad} \ge 0.99 P$ at any depth, it is set equal to $0.99P$ to avoid
negative $P_{\rm Gas}$ values that would lead to negative $\rho$ values.  This fix is necessary for modelling
with $T_{\rm Eff}$ values above $\approx 12000$ K for dwarfs.  Hydrostat.hydrostatic() returns a
$4 \times 50$ array in which rows 0 and 1 hold linear and logarithmic $P_{\rm Gas}$ values, and rows 2 and 3  hold
linear and logarithmic $P_{\rm Rad}$ values.

\paragraph{}

5) Method state() in Java class State (grayfox.State.state(), ported to JavaScript function state()) evaluates the equation of state (EOS) under
the assumption of an ideal gas, and computes the $\log\rho(\log\tau_{\rm Ros})$ structure.  I have assumed
a mean molecular weight, $\mu$, of 1.0, between the extreme values of 0.62 and 1.3 for gas of solar composition
that is completely ionized and completely neutral, respectively.

\paragraph{}

6) Method depthScale() in Java class DepthScale (grayfox.DepthScale.depthScale(), ported to JavaScript function depthScale()) computes the geometric
depth scale, $z(\log\tau_{\rm Ros})$ corresponding to the $\tau_{\rm Ros}$ scale by numerically evaluating
the finite difference definition of ${\rm d}\tau$ in terms of $\kappa$,
${\rm d}\tau_{\rm Ros}(z) = \rho(z) \kappa_{\rm Ros}(z) {\rm d}z$.  I set the top of the atmosphere to have a depth of 0 cm
and evaluate the $z$-scale with Euler's method.  The $z$-scale
is very sensitive to the upper boundary condition because it depends on the value of $\rho$ through the
definition of ${\rm d}\tau$ in terms of $\kappa$.  The value of $\rho$ computed in Step 5) is artificially
small at the upper boundary because of the necessarily low value of $P$ chosen for the upper boundary
value in Step 4).  Therefore, I compute the first two values of $z$ at the top of the atmosphere
using the value of $\rho$ at the third depth point inward.

\paragraph{}

7) Method formalSoln() in Java class FormalSoln (grayfox.FormalSoln.formalSoln(), ported to JavaScript function formalSoln()) computes the outgoing
logarithmic monochromatic emergent surface specific intensity, $\log I_\lambda(\tau=0, \cos\theta)$ {\it
vs} $\cos\theta$ for $0 \le \cos\theta \le 1.0$, by evaluating the formal solution of the radiative
transfer equation for a $1D$ plane-parallel semi-infinite geometry.  The formal solution is numerically
evaluated on the $\log\tau_{\rm Ros}$ scale with the extended rectangle rule.  FormalSoln.formalSoln()
calls method planck() in Java class Planck (Planck.planck()) to evaluate the Planck function, $B_\lambda$,
with the radiation temperature set equal to the local kinetic temperature, for use as the LTE source
function, $S_\lambda$, {\it ie.}
$S_\lambda (\log\tau_{\rm Ros}) = B_\lambda (\log\tau_{\rm Ros}) = B_\lambda (T_{\rm Rad} = T_{\rm Kin}(\log\tau_{\rm Ros}), \lambda)$.
The $\log I_\lambda(\cos\theta)$
distribution is evaluated at a set of nine $\theta$ values for zero and the the positive half ($[0, 
1]$) of the 17-point Gauss-Legendre quadrature on the domain $[-1, 1]$, sufficient to exhibit the rapid decline in
$I_\lambda(\cos\theta)$
near the limb.  This was re-scaled to sample the $\theta$
range $[0, \pi/2]$ RAD.  The $I_\lambda(\lambda)$ distribution is evaluated at a set of 20 $\lambda$ values
distributed evenly in $\log\lambda$ from 200 nm in the UV to 2000 nm in the IR.

\paragraph{}

8) The method flux() in Java class grayfox.Flux (grayfox.Flux.flux(), ported to JavaScript function flux()) computes the emergent
monochromatic surface flux, $F_\lambda(\tau=0)$,
by Gaussian quadrature for my $\lambda$ set.  At this point I check the quality of the calculation by
numerically integrating $F_\lambda(\tau=0)$ in $\lambda$ with the extended rectangle rule to produce
$F_{\rm Bol}$, and recovering a value for $T_{\rm Eff}$ from $(F_{\rm Bol}/\sigma)^{1/4}$, where $\sigma$ is the Stefan-Boltzmann constant.

\subsection{Spectral line profile calculation}

The following Java classes contain the methods that handle each module of the spectral line synthesis
problem and are contained in Java package grayfox and are also ultimately launched by invocation of
GrayFox.main().

\paragraph{}

1) Method voigt() in Java class LineGrid (grayfox.LineGrid.voigt(), ported to JavaScript function voigt()) establishes a grid of $\Delta\lambda$
points around line center, $\lambda_{\rm 0}$, that samples a representative spectral line, and computes
the line profile, $\phi_\lambda$.  The Doppler width, $\Delta\lambda_{\rm D}$, is computed accounting
for both the thermal Doppler broadening at a temperature equal to $T_{\rm Eff}$ for an absorbing particle
of the mass provided as an input parameter, and the turbulent Doppler broadening for the input value
of $\xi_{\rm T}$.  This establishes a depth-independent scale for the Voigt $v$ parameter,
$v = \Delta\lambda /\Delta\lambda_{\rm D}$.  I distribute five $v$ points that sample the positive range ({\it ie.} ``red''
half of the line profile) uniformly from 0 ($\lambda = \lambda_{\rm 0}$) to 5 to sample the Gaussian
core of the $\phi_\lambda$ function.  I distribute five more $v$ points uniformly in $\log v$ from
$v \approx > 0$ to $\log v = 1.5$ to sample the positive Lorentzian wing of the $\phi_\lambda$ function,
for a total of 10 zero or positive $v$ points.  The Voigt $a$ parameter is computed with a van der Waals-type
broadening parameter, $\gamma$, that is scaled depth-wise with $T_{\rm Kin}$ and $P_{\rm Gas}$ according
to the formula in \citet{rutten} with respect to a solar value of $\log\gamma \approx 9$ $\log$ s$^{-1}$ at
$\tau_{\rm Ros}=1$, then re-scaled again with the input value of the damping enhancement parameter.  The Voigt function,
$H(a, v)$ is approximated as $e^{-v^2} + a/(\pi^{1/2} v^2)$.  For $v$ values $\le$ 1.0, I compute
$\phi_\lambda$ as a pure Gaussian, $H(a, v) \approx e^{-v^2}$.  This piece-wise evaluation is necessary
because the Lorentzian term, $ a/(\pi^{1/2} v^2)$, is singular at line center ($v=0$).  These choices
for the numbers of core and wing $v$ points, and the ranges for the core and wing $v$ distributions, was
found by trial-and-error to give rise to smooth line profiles all along the curve-of-growth (COG), including
saturated lines with distinct Lorentzian wings.  The normalized $\phi_\lambda$ function is found
from $H(a, v) / (\pi^{1/2} \Delta\lambda_{\rm D}$).  The blue half ($v < 0$) of the line profile is filled
in by reflection of $\phi_\lambda(\lambda)$ about $\lambda_{\rm 0}$.

\paragraph{}

2) Method levelPops() in Java class LevelPops (grayfox.LevelPops.levelPops(), ported to JavaScript function levelPops()) computes number
densities of particles in the lower level of the representative {\it b-b} transition for the
line extinction calculation, as well as in
the upper transition level and in the ground states of the neutral and ionized stages for pedagogical
display of the excitation and ionization equilibrium.  The input number density of absorbers, $N$, is scaled by
$\rho(\log\tau_{\rm Ros})/\rho(\log\tau_{\rm Ros} = 1)$ to produce a depth-dependent number density,
$N(\tau_{\rm Ros})$ that reflects the vertical density
variation of the atmosphere. The $N (\log\tau_{\rm Ros})$ distribution is scaled by either $(1 + y)$
or $(y/(1 + y))$ depending on whether the absorbing species is of the neutral stage
($\chi_{\rm l} < \chi_{\rm I}$) or the singly ionized stage ($\chi_{\rm l} > \chi_{\rm I}$), respectively, where
$y = N_{\rm II}/N_{\rm I}$ and is computed from the Saha equation.  The depth-dependent value of the free
electron density, $N_{\rm e}(\log\tau_{\rm Ros})$, in the Saha equation is estimated from the relation
$\log N_{\rm e} (\log\tau_{\rm Ros}) \approx -10.6 - \log k + 0.5\log T_{\rm Kin}(\log\tau_{\rm Ros})$,
suitable for early-type dwarfs \citep{turner}.  The $N (\log\tau_{\rm Ros})$ distribution is then re-scaled by
$e^{-\chi_l /kT_{\rm Kin}(\tau_{\rm Ros}})$, where $k$ is Boltzmann's’s constant, to produce the depth-wise number density of
absorbers in $E$-level $l$, $N_{\rm l}(\tau_{\rm Ros})$.  Our approximate
treatment of excitation and ionization equilibrium assumes that the statistical weight of the level,
$g_{\rm l}$, and the partition functions for the neutral and singly ionized species are all unity.

\paragraph{}

3) Method lineKap() in Java class LineKappa (grayfox.LineKappa.lineKap(), ported to JavaScript function lineKap()) computes the depth-dependent
monochromatic line mass extinction co-efficient, $\kappa_\lambda^l(\log\tau_{\rm Ros}, \lambda)$, from
the normalized $\phi_\lambda$ profile computed in Step 1), the energy level population, $N_{\rm l}(\tau_{\rm Ros})$,
from Step 2), and the input value of $f_{\rm lu}$.
I account for the LTE stimulated emission correction in my computation of $\kappa_\lambda^l(\log\tau_{\rm Ros})$.

\paragraph{}

4) Method tauLambda() in Java class LineTau2 (grayfox.LineTau2.tauLambda(), ported to JavaScript function tauLambda()) computes the  monochromatic
line optical depth scale, $\tau_\lambda^l(\log\tau_{\rm Ros}, \lambda)$, from the
$\kappa_\lambda^l(\log\tau_{\rm Ros}, \lambda)$ distribution computed in Step 3).  Methods FormalSoln.formalSoln() and Flux.flux() (see
Step 7 in the atmospheric model and SED procedure above) are invoked with
$\kappa_\lambda(\log\tau_{\rm Ros}) = \kappa_\lambda^l(\log\tau_{\rm Ros}) + \kappa_{\rm Ros}(\log\tau_{\rm Ros})$ to compute the line
$\log I_\lambda(\tau=0, \cos\theta)$ and $F_\lambda(\tau=0)$ distributions, respectively.  The value
of $\log I_\lambda(\tau=0, \cos\theta)$ and $F_\lambda(\tau=0)$ is also computed at
$\lambda = \lambda_{\rm 0}$ with $\kappa_\lambda(\log\tau_{\rm Ros}) =  \kappa_{\rm Ros}(\log\tau_{\rm Ros})$ for use in producing
continuum normalized line profiles, $I_\lambda(\tau=0, \cos\theta)/ I^{\rm c}_{\lambda 0}(\tau=0, \cos\theta)$
and $F_\lambda(\tau=0)/ F^{\rm c}_{\lambda 0}(\tau=0)$.

\paragraph{}

\subsection{Additional observables}

The following Java classes produce additional standard observables of pedagogical significance:

\paragraph{}

Method UBVRI() in Java class Photometry (grayfox.Photometry.UBVRI(), ported to JavaScript function UBVRI()) produces five calibrated photometric
colour indices in the Johnson-Cousins $UBV(RI)_{\rm C}$ system, $U-B$, $B-V$, $V-R$, $V-I$, and $R-I$.
 Our colours will necessarily be crude, particularly those involving the $U$ and $B$ filters, because
my gray atmosphere SED is unblanketed.  Therefore, I do not use realistic filter transmission functions,
$T(\lambda)$.  Instead, I compute the flux in each photometric band with a three-point approximation
that samples $T(\lambda)$ at band center, $\lambda_{\rm 0}$ and at the red and blue half-power points,
$\lambda_{\rm 1/2, Red}$ and  $\lambda_{\rm 1/2, Blue}$.  I calibrate the colour indices with a single
point calibration to Vega (A0 V, $\alpha$ Lyr, HR 7001, HD 172167) using raw colours computed with a GrayFox
model with input parameters of $T_{\rm Eff} = 9550$ K and $\log g = 3.95$ for Vega \citep{castelli}.
 Method eqWidth() in Java class EqWidth (grayfox.EqWidth.eqWidth(), ported to JavaScript function eqWidth()) computes the equivalent
width, $W_\lambda$, of
the continuum normalize flux spectral line, $F_\lambda(\tau=0)/F^{\rm c}_{\lambda 0}(\tau=0)$, using the
extended rectangle rule.

\subsection{Visualization }

  Neither JavaScript nor HTML are a plotting or a graphics package in {\it any} sense.  Unlike plotting packages such as IDL, gnuplot, {\it etc.},
JavaScript and HTML do not have intrinsic functionality for converting data coordinates to device coordinates, scaling and graduating axes with canonical tick values,
recognizing arrays as being ordinates or abscissae of plots, or with interpolating between adjacent
data points to produce line graphs.  Plotting package functionality
must be manually programmed starting from the primitive ability of JavaScript to create HTML elements, which are nothing more than rectangular
areas oriented square to the browser window axes, and to set their style attributes, such as their top- and left-margin offsets in absolute device
pixels.  Every graphical element comprising the plotted output, including each individual plot symbol and tick-mark, is an individual HTML $<$div$>$
element with its top- and left-margin device pixel coordinates computed from the output of the modelling calculation.
As a result, almost half of the 4500 lines of JavaScript code are devoted to producing the graphical output.  In my judgment, this lack of intrinsic
plotting functionality is a larger barrier to scientific modelling and visualisation in JavaScript and HTML than is any limitation in
JavaScript's ability to handle scientific computing algorithms.

  \section{Pedagogical applications \label{sApps}}

 It is worth noting that the GrayStar GUI is an HTML Web page like any other, and thus the usual methods for managing the display of content, and for capturing
textual content, are available: A presenter can enhance clarity by zooming, isolate areas of interest by re-sizing the browser, and show direct comparison
of output from runs with different parameters by launching multiple instances of the browser, or by using multiple browser tabs, each running 
GrayStar, and a student can capture textual output such as colour indices, $W_{\lambda}$ values, {\it etc.}, by cutting and pasting to a 
common spreadsheet program for analysis and plotting.  In particular, because GrayStar's graphical output consists of 
HTML instructions to the browser's rendering engine rather than pixelated bitmap information (such as that encoded in jpegs, gifs, {\it etc.}),
the graphics are scale-invariant and remain sharp at high zoom factors, which is an important consideration when presenting in large
classrooms.
Particular pedagogical applications are are discussed in Paper I.

\paragraph{}

At the advanced undergraduate or graduate level, students can be asked to modify and add to the source code itself.  To modify and run the Java development version,
instructors and students need to download and install JDK 1.8 or later, and version 8.0 or later of the NetBeans IDE,
both available free of charge from Oracle's WWW site.
To trouble-shoot the JavaScript deployment version with diagnostic print statements (console.log()), the ``Developer tools'' accessible 
from the browser menu must be enabled, and the ``console'' selected. 

%
%
%

\section{Conclusions}

 GrayStar allows real-time exploration and investigation of stellar atmospheric structure and spectral
line profiles with ``on-demand'' parameters suitable for classroom demonstration and student laboratory 
assignments.  The JavaScript and HTML code is robustly platform-independent across all common types of
university-supplied and student-owned computers.  GrayStar allows pedagogical demonstration of 
most, if not all, major topics in the undergraduate astrophysics curriculum that are related to 
stellar atmospheres and spectra.  A local installation of GrayStar can be embedded in the local
pedagogical framework by editing the pedagogical links. 
 In addition to making GrayStar available for local installation and use by the astronomy teaching community,
I also encourage active development and adaptation of the code.  

\paragraph{}

  JavaScript, a language that can be interpreted by any common Web browser and executed on any common commodity 
personal computer for which a modern browser is available, is sophisticated enough as a programming language 
to allow development of scientific simulations at at least a pedagogically useful level of realism.  The ability of
JavaScript to manipulate HTML documents allows the results of simulations to be visualised in the Web browser.   
Commodity computers are now powerful enough to execute such JavaScript simulations instantaneously.   This allows 
``toy'' models of natural and physical systems to be simulated and visualised in a way that allows for 
pedagogical experimentation, classroom demonstration, and exploration of parameter space with no requirement
of computational or visualisation skills on the part of the instructor or student, and with no need
to install special purpose software.  Codes can be developed in Java, thus taking advantage of the powerful and mature 
developer support framework for Java, including IDEs, then straightforwardly ported to JavaScript.  

\paragraph{}

The biggest dis-incentive to scientific programming and visualisation with JavaScript and HTML is the need to 
manually emulate the functionality of plotting packages ({\it eg.} IDL) starting from the primitive ability
to place a rectangle of given dimensions at a given location in the browser window in absolute device coordinates.
However, the code in the graphical output section of GrayStar may be taken as a template for how this problem can be 
addressed, and adapted to other uses.

\paragraph{}

This opens 
up an entire vista of computational pedagogical possibilities, and the pedagogical stellar atmospheric and spectral line
modelling described here is only one example.  For example, within the field of stellar astrophysics, a similar approach could be taken for 
pedagogical simulation of stellar interior structure in the polytrope approximation, allowing classroom demonstration and
student laboratory investigation of the dependence of stellar structure and related observables on various
independent parameters - a ``stellar interior with knobs''. 
I stress again the broader significance that common PCs running common OSes are now powerful enough, and common Web-browsers are now sophisticated
enough interpreters, that the the realm of pedagogically useful scientific programming is now accessible in a framework that is free,
common, and allows both the application and its source code to be immediately shared over the Web.

\paragraph{}



\acknowledgments

\clearpage



\clearpage







\end{document}